\documentclass[aps,prr,reprint,twocolumn,superscriptaddress,floatfix,nofootinbib,longbibliography]{revtex4-1}
\usepackage{amsmath,amssymb,color,comment,physics,mathtools}
\usepackage[makeroom]{cancel}
\usepackage[caption=false]{subfig}
\usepackage{mathrsfs} 
\usepackage{color} 
\usepackage{graphicx}
\usepackage{subfig}
\usepackage[countmax]{subfloat}
\usepackage[english]{babel}
\usepackage[bookmarks=true,colorlinks,linkcolor=OrangeRed,urlcolor=NavyBlue,citecolor=RoyalBlue]{hyperref}
\usepackage[dvipsnames]{xcolor}
\usepackage{ulem} 
\usepackage{braket}
 \usepackage{dsfont}

\newcommand{\be}{\begin{eqnarray}}
\newcommand{\ee}{\end{eqnarray}}

\begin{document}

\title{ Dominant Reaction Pathways by Quantum Computing}
\author{Philipp Hauke} 
\affiliation{INO-CNR BEC Center and Department of Physics, University of Trento, Via Sommarive 14, I-38123 Trento, Italy}
\author{Giovanni Mattiotti} 
\affiliation{Department of Physics, University of Trento, Via Sommarive 14, I-38123 Trento, Italy}
\author{Pietro Faccioli}
\affiliation{Department of Physics, University of Trento, Via Sommarive 14, I-38123 Trento, Italy}
\affiliation{INFN-TIFPA, Via Sommarive 14, I-38123 Trento, Italy}

\begin{abstract}

Characterizing thermally activated transitions in high-dimensional rugged energy surfaces is a very challenging task for classical computers.
Here, we develop a quantum annealing scheme to solve this problem. 
First, the task of finding the most probable transition paths in configuration space is reduced to a shortest-path problem defined on a suitable  weighted  graph.  Next, this optimization problem is mapped into finding the ground state of a generalized Ising model. A finite-size scaling analysis suggests this task may be solvable efficiently by a quantum annealing machine. Our approach leverages on the quantized nature of qubits to describe transitions between different system's configurations. Since it does not involve any lattice space discretization, it paves the way towards future biophysical applications of quantum computing based on realistic  all-atom models. 
\end{abstract}

\date{\today}
\maketitle

Molecular Dynamics (MD) provides a sound theoretical framework to investigate the structural dynamics of macromolecular systems. However, this scheme is  computationally inefficient when applied to  thermally activated processes.
To overcome this limitation,  much effort has been put  over the last two decades toward devising enhanced sampling algorithms \cite{enhanced}. 
In particular, an exponential speed-up of the computational efficiency can be obtained by introducing specific biasing forces which promote the escape rate from metastable minima~\cite{BFA, METAD, AMD}. 
However,  methods based on this scheme typically require some prior knowledge about the reaction coordinate or the system's slowest  Collective Variables (CVs). Unfortunately, the identification of these variables is in general a very challenging  task, and  a sub-optimal choice may hamper the convergence or introduce systematic errors.  On the other hand, enhanced sampling methods  which do not involve biasing forces \cite{TPS, MSM, milestoning}
are significantly more computationally expensive.

During the last several years, quantum computing machines  have grown exponentially  both in size and performance, to a point that  it is now realistic to foresee the onset of quantum supremacy in key computational problems \cite{Arute2019}. It is therefore both important and timely to address the question whether quantum computation can be employed to identify statistically relevant transition pathways in high-dimensional rugged energy surfaces.

While significant progress has been made on designing algorithms for quantum annealers \cite{Das2005,Das2008,Albash2018,Venegas2018,Hauke2020} that specifically tackle quantum chemistry  applications \cite{Streif2018,Li2018,Genin2019,Cao2019,Matsuura2020,Outeiral2020},  only a few applications  to classical molecular sampling problems have been reported to date  \cite{QMD1, QMD2}. Arguably, the key limiting factor is the fact that quantum machines are best suited to tackle  discrete  problems. For this reason, to the best of our knowledge, all the  proposed quantum computing algorithms  for sampling and energy optimization of classical molecular structures rely on simplified lattice models. While  these  models  have provided valuable insight into the general statistical mechanical properties of biopolymers \cite{lattice_models}, the lack of structural and chemical detail hampers their applicability to  realistic biophysical systems. 

In this work, we develop a rigorous approach to finding the most statistically relevant transition paths in a thermally activated conformational reaction, using a quantum computing machine. Our method  does not require lattice discretization. Thus, it is  in principle applicable to realistic molecular models, with atomic resolution. Although the resources available in present-day quantum computers limit immediate implementations to rather small benchmark problems, our work points to a paradigmatically new way to tackle problems from computational biophysics.

The basic idea is to first resort to classical computing  to generate large data sets of molecular  conformations, mostly concentrated in the transition region. Unlike in Markov State Models \cite{MSM} or in the milestoning approach \cite{milestoning},  this set of molecular configurations  does not need to be generated with a realistic dynamics, nor to sample a physically meaningful distribution. The only request is to explore the relevant region of configuration space. For example, one could use plain MD at high temperature,  machine-learning schemes for uncharted manifold learning \cite{Covino}, or biased dynamics or combinations of these methods. 

The next step is to assign {\it a posteriori} the correct relative statistical weight to all the reactive trajectories  that can be drawn  by connecting the configurations in this  sparse dataset. 
Indeed, using the so-called Dominant Reaction Pathways (DRP) formalism \cite{DRP1} the most probable transition pathways connecting given initial and final configurations can be  rigorously obtained from a least-action principle. 
After representing configuration space with a sparse set of molecular conformations,  the DRP variational principle becomes equivalent to a shortest path  problem formulated on a discrete weighted graph, a problem which can be tackled by quantum annealing. The key difference to other proposed quantum computing approaches for molecular sampling \cite{QMD1, QMD2} is that the discretization is performed at the level of the system's configuration space and that the points in this configuration space are defined in continuum space and are generated using a classical computer.

{\it Dominant Reaction Pathways.}  Let us begin by briefly reviewing the DRP formalism. The starting assumption is that the system's structural dynamics can be modeled by a set of  over-damped Langevin equations:
\be\label{Lang}
\dot {\bf q}_i = -\frac{D}{k_BT} \nabla_i U({\bf q}) + \eta_i(t),\qquad i=1, \ldots N. 
\ee
 ${\bf q}_i$  denotes the position of the $i-$th particle in the system, and   $D$ is the  diffusion coefficient (here chosen for simplicity to be the same for all particles).  $U(Q)$ is the molecular potential energy and  $Q=({\bf q}_1, \ldots {\bf q}_N)$ is the $3N-$dimensional coordinate in the system's configuration space, while $\eta_i(t)$ is a delta-correlated white noise  obeying the fluctuation-dissipation relationship $\langle \eta_i(t) \eta_i(0) \rangle = 2 D \delta_{ij} \delta(t) $.
The over-damped limit is appropriate to analyze the  dynamics of  biomolecules, with a time-resolution on the order of picoseconds or lower.
 \begin{figure}[t!]
	\includegraphics[width=.5\textwidth]{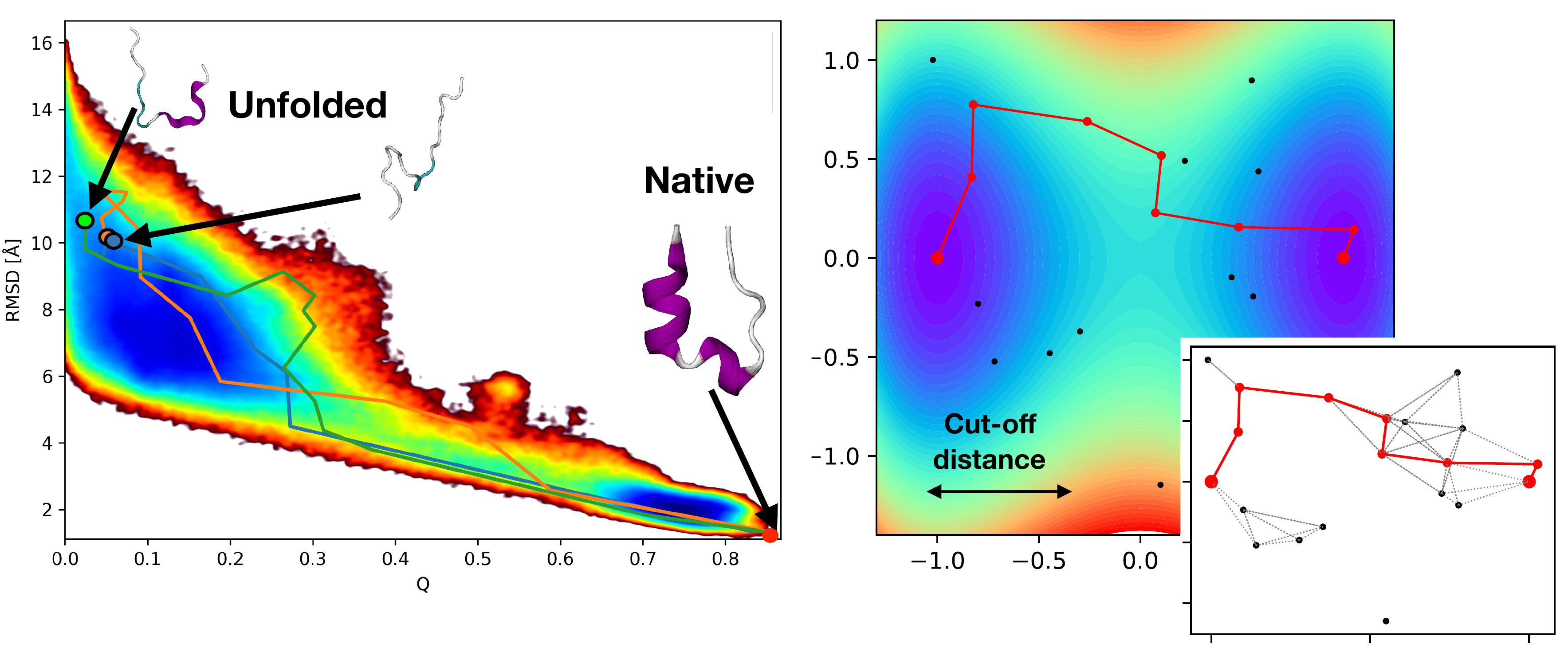}
	\caption{Illustrative applications of our reaction pathway calculations in prototypical systems. Left panel: The heat-map in the background shows the potential of mean-force estimated from plain MD as a function of  the RMSD to the crystal native structure of protein Trp-Csage and the fraction of native contacts $Q$.  The solid lines correspond to dominant reaction pathways calculated using the Dijkstra algorithm from three different unfolded configurations in a graph with 2180 vertices and $4\times 10^4$ directed edges.  Each dominant trajectory consists of about 20 frames.
	Right Panel:  Calculation of dominant reaction path in a two dimensional double-well potential, using a hybrid classical/quantum annealing approach on a D-Wave machine.}
	\label{fig:concept}
\end{figure}

The probability of finding the system in a conformation
$Q_f$ at time $t$, conditioned to be at the conformation $Q_0$ at  the initial time $t_0$ can be expressed as a path integral, 
\be\label{PI}
 P(Q_f, t| Q_0, t_0) = e^{-\frac{(U(Q_f)-U(Q_i))}{2 k_BT}}\,\int_{Q_0}^{Q_f} \mathcal{D} Q \,e^{-S_{\mathrm{eff}}[Q]}, 
 \ee
where $S_{\mathrm{eff}}[Q]$ is called the effective action and reads
\be\label{Seff}
S_{\mathrm{eff}}[Q] = \int^{t}_{t_0} d\tau \left(\frac{\dot Q^2}{4 D} + V_{\mathrm{eff}}[Q(\tau)] \right),
\ee
while $V_{\mathrm{eff}}(Q) = \frac{D}{4 (k_BT)^2} \left( (\nabla U(Q))^2- 2 k_BT \nabla^2U(Q)\right)$.
 
The DRP approach focuses on the most probable reactive trajectories and is based on the saddle-point approximation of the functional integral (\ref{PI}). The functional  minima of the effective action $S_{\mathrm{eff}}[Q]$, are called dominant reaction  paths and obey Newton-type  equations of motion:   $2D \ddot {\bf q}_i = \nabla_i V_{\mathrm{eff}}(Q)$. This implies the conservation of the so-called effective energy $E_{\mathrm{eff}} = \frac{\dot Q^2}{4D} -V_{\mathrm{eff}}(Q)$ along the reaction paths.  
Using this property,  it is possible to show that the saddle-points of the functional integral (\ref{PI}) coincide with the minima of the Hamilton--Jacobi (HJ) functional \cite{DRP0, DRP1}
\be\label{SHJ}
S_{\mathrm{HJ}}[Q] = \int_{Q_0}^{Q_f} dl \sqrt{\frac{1}{D} (V_{\mathrm{eff}}[Q(l)] + E_{\mathrm{eff}})}, 
\ee
where $dl=\sqrt{dQ^2}$ is  an infinitesimal conformational
displacement along the dominant path.  
The effective energy parameter  $E_{\mathrm{eff}}$ implicitly sets the time interval $t-t_0$ entering Eq. (\ref{PI}), through the equation $t-t_0= \int_{Q_0}^{Q_f} dl \{4D (V_{\mathrm{eff}}[Q(l)]+ E_{\mathrm{eff}}) \}^{-1/2}$. For this time-scale to be on the order of the typical transition path time,  the effective energy must be  chosen $E_{\mathrm{eff}}\sim -V_{\mathrm{eff}}(Q_i)$ ~\cite{DRP2, hMD}.

In  molecular simulations,  the HJ formulation of the saddle-point condition leads  to  a dramatic computational advantage with respect to the standard  (i.e.\ time-dependent) formulation. This is because the  root-mean-square distance (RMSD) between kinetically distant molecular configurations is less than two orders of magnitude larger than the atomic size (see e.g. the example in the left panel of Fig.~1). As a consequence, only a few tens of configurations are usually sufficient for an accurate representation of the line integral  in the HJ functional  (\ref{SHJ}). In contrast,  to accurately discretize the time integral in the effective action (\ref{Seff})  one would need to use time steps $dt$ on the order of fs, while the typical transition path times for complex macromolecular transitions (such as e.g.\ protein folding) are  about 9 orders of magnitude larger. Thus, using the HJ formalism it is possible to assign correct statistical weight to the paths obtained by connecting a string of structurally close configurations, even though the typical time it would take the system to diffuse between them may be very long. 
To sample the transition path ensemble, many independent minimizations of the HJ functional must be performed, corresponding to different choices of initial and final configurations, in the reactant and product states. 

{\it Shortest-path problem formulation.} The HJ saddle-point condition can be  mapped into a shortest-path problem  on a discrete graph. To this end, we assign a vertex in the graph (labelled $i,j=1\dots |\mathcal{V}|$) to each molecular configuration in the previously generated dataset and identify the initial conformation $Q_0$ with the source $\mathrm{s}$ and the final conformation $Q_f$ with the target $\mathrm{t}$.  
Next, we define a set $\mathcal{E}$ of directed edges $(ij)$ that connect vertices $i$ and $j$. Each directed edge is assigned a weight  (or `cost') 
\be\label{SHJ1}
w_{(ij)} &=&  \frac{ \Delta l_{(ij)}}{2 \sqrt{D}} \left( L_i+L_j \right)
\ee
with $L_i= \sqrt{V_{\mathrm{eff}}(Q_i)+E_{\mathrm{eff}}}$ and $\Delta l_{(ij)}=\sqrt{(Q_i-Q_j)^2}$.  In this notation, the target function to minimize on the network is $S_{\mathrm{HJ}} = \sum_{k=1}^M w_{(i_k  i_{k+1})}$, with $i_1=\mathrm{s}$ and $i_M=\mathrm{t}$. 
To ensure that the HJ functional is correctly represented by discrete paths on the network, we take only edges into account with a cost below a given cutoff $W_\mathrm{c}$. We note that this condition can drastically reduce the complexity of our optimization problem, as for a judicious choice of $W_\mathrm{c}$ the number of edges will become proportional to the number of vertices $|\mathcal{V}|$ rather than $|\mathcal{V}|^2$.

To summarize, so far we have mapped the original problem of finding the dominant reaction pathways in a continuous energy landscape into a shortest path problem in the weighted graph ($\mathcal{V}$, $\mathcal{E}$), that  can be efficiently tackled  with discrete optimization methods.

In the Supplementary Material (SM) \cite{SM}, we illustrate this approach  in a  prototypical two-dimensional toy system. Here, we focus  on the first realistic application to a small-scale  bio-molecular transition. In particular, we study the folding of the Trp-Cage protein  (pdb code: 2JOF, see Fig.~1). Since this  is one of the smallest and fastest folding mini-proteins,  its folding mechanism has been characterized  by plain MD, using  a special purpose supercomputer \cite{howfastfoldingproteinsfold}.

As a first step, we built  the graph  representation of  the folding kinetics, by  assigning a vertex  to each of 2180 independent protein configurations, generated with the so-called Self Consistent Path Sampling (SCPS) enhanced sampling algorithm \cite{SCPS, HET-s}, using the same all-atom force field as adopted in  Ref.~\cite{howfastfoldingproteinsfold}. Then,   $4\times10^4$ directed weights $w_{ij}$ on the graph were computed according to Eq.~(5) (further details are provided in the SM \cite{SM}). The number of vertices and edges were chosen to ensure  connectivity between nearest-neighbor configurations with RMSD $\le $3 \AA.  Finally, most probable  pathways connecting  different unfolded configurations to the same (crystal) native structure were calculated by finding the shortest distance on the graph, using the Dijkstra algorithm  \cite{Dijkstra}.

The results of DRP and plain MD simulations are compared in Fig.~1, projected on the plane selected by two commonly adopted CVs. The calculated trajectories travel along the region of low free-energy,  estimated by MD. Each path consists of about $ 20$ frames and connects configurations harvested from different SCPS trajectories. This fact is crucial:  It shows that our algorithm is efficiently combining the  information provided by different molecular datasets.

{\it Finding the Shortest-Path by Quantum Computing}. Let us finally tackle the problem of formulating the shortest path problem in a way amenable to  global optimization through a quantum annealing machine \cite{Das2005}.  To this end, we consider the linear programming formulation of the shortest path problem:
We introduce the binary variables $x_{(ij)}=0,1$, located at edges $(ij)$. $x_{(ij)}=1$ means the edge $(ij)$ is part of the shortest path, while $x_{(ij)}=0$ means it is not. We emphasize that in our directed graph, we distinguish between $(ij)$ and $(ji)$.  The task of finding the shortest path is then translated into minimizing the target function $\sum_{(ij) \in \mathcal{E}} w_{(ij)} x_{(ij)}$ subject to the constraints
\begin{align}
\label{eq:constraints}
	&\forall i: \sum_{j\in \mathcal{V}} x_{(ij)} - \sum_{j\in \mathcal{V}} x_{(ji)} = \begin{cases}1, &\text{if }i=\mathrm{s};\\ -1, &\text{if }i=\mathrm{t};\\ 0, &\text{ otherwise.}\end{cases}
\end{align}
These constraints  represent a conservation of flux: since we are searching for a connected path, at each vertex the number of incoming edges must  equal  the number of outgoing edges. Exceptions are $\mathrm{s}$, where the number of outgoing edges is larger by one than the number of incoming edges, and $\mathrm{t}$, where the converse is true. 

These can be cast into the form of a quadratic function 
\be
&&H_A = A \left\{ 
				\left[\sum_{j\in \mathcal{V}} \left(x_{(\mathrm{s}j)}-x_{(j\mathrm{s})}\right)-1\right]^2 +	\right.	 \nonumber\\
				&&\left. \left[\sum_{j\in \mathcal{V}} \left(x_{(\mathrm{t}j)} - x_{(j\mathrm{t})}\right) +1 \right]^2	+ \sum_{\stackrel{i \in \mathcal{V}}{ i \neq \mathrm{s}, \mathrm{t}}}\left[\sum_{j\in \mathcal{V}} \left(x_{(ij)} - x_{(ji)}\right) \right]^2			\right\} \nonumber\\
\ee
Note that in the above terms, the summation over $j$ is restricted to the vertices connected to $\mathrm{s}$, $\mathrm{t}$, and $i$, respectively. Defining in addition the cost function to the HJ functional
\be
H_B = B \sum_{(ij) \in \mathcal{E}} w_{(ij)} x_{(ij)}, 
\ee
the shortest path problem can be reformulated as: Find the set of binary variables $x_{(ij)}$ that minimize the classical cost function $H = H_A + H_B$. 
Note that, in order for $H_A$ to be a hard constraint, we require the ratio between coupling constants $A/B$ to be sufficiently large. A rigorous condition is $A/B>|\mathcal{E}|\, W_\mathrm{c}$, where $|\mathcal{E}|$ is the number of edges, but in practice smaller values often suffice.

In order to solve the problem with a quantum processing device, we re-interpret the classical cost function $H$ as a quantum mechanical Hamiltonian described by $z$-Pauli-components of spins $1/2$, defined by  $\sigma_{(ij)}^z=2 x_{(ij)}-1$. These spins $1/2$ are to be encoded in the qubits of the quantum annealer. The resulting generalized Ising Hamiltonian is explicitly reported in the SM \cite{SM}.

In quantum annealing, the problem of finding the minimum energy configuration of $H$ is solved in the following way \cite{Das2005,Das2008,Albash2018,Venegas2018,Hauke2020}. The qubits are initialized in the ground state of an easily solvable Hamiltonian that does not commute with $H$, say $H_\mathrm{in}=-h\sum_{(ij)}\sigma_{(ij)}^x$. Then, the qubit system is subjected to a time-dependent Hamiltonian $a(t) H_\mathrm{in} + b(t) H$, with scheduling functions $a(t)$ and $b(t)$. These are chosen such that initially $a(0)=1$ and $b(0)=0$, while at the end of the protocol at $t_\mathrm{sweep}$ one has $a(t_\mathrm{sweep})=0$ and $b(t_\mathrm{sweep})=1$. In this way, the Hamiltonian is transformed from the simple $H_\mathrm{in}$ to the complex problem Hamiltonian $H$. The adiabatic theorem implies that if the sweep is performed sufficiently slowly as compared to the minimal energy gap encountered, then the system remains in its instantaneous ground state.  That is, the final state at the end of the protocol is the ground-state of $H$, i.e.\ the solution to the shortest-path problem. Note that in current practical devices there is considerable coupling to the environment, such that the sweep is not fully coherent and higher levels can be populated. This deteriorates the success probability and makes it difficult to derive rigorous proofs about any quantum advantage. Despite these limitations, quantum annealing has been used successfully for a large number of problems, including many NP-complete as well as applications ranging from search-engine ranking over machine learning to quantum chemistry, and various ways forward to improve the performance of quantum annealing devices have been identified; see Refs.~\cite{Albash2018,Venegas2018,Hauke2020} for recent reviews. 

\begin{figure}
	\includegraphics[width=.45\textwidth]{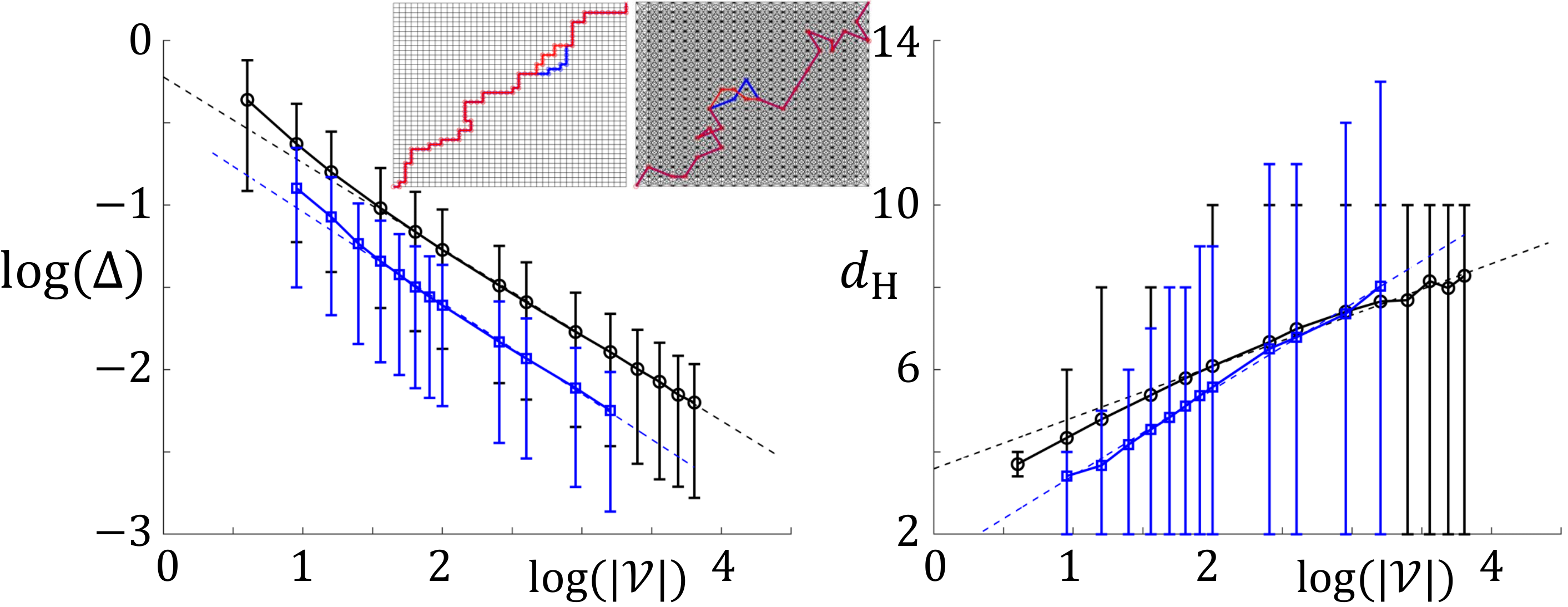}
	\caption{
			Scaling versus number of vertices $|\mathcal{V}|$ of average energy gap $\Delta$ between the two lowest classical solutions (left) and corresponding Hamming distance $d_\mathrm{H}$ (right). 
			Data is for a square lattice with edges between nearest neighbor (NN, black circles) and up to next-next-to-nearest neighbor vertices (NNNN, blue squares). The insets show example paths on these lattices. 
			Error bars are $67\%$ confidence intervals and dashed lines are linear fits (see SM \cite{SM}). 
			The approximately polynomial decrease of gap and logarithmic increase of Hamming distance suggest the efficiency of the problem for a quantum annealer. 
		}
	\label{fig:scaling}
\end{figure}

As for the problem at hand, it is known that the shortest-path problem can be solved in polynomial time on classical computers \cite{runningtime}. One may then wonder if the quantum algorithm can compete with this scaling. 
To shed light on this issue, we performed a careful finite-size scaling of our target problem $H$ using a classical linear program, from which we can estimate the order of the gap closing during the annealing sweep \cite{Altshuler2010} (for details, see SM \cite{SM}). We focus on regular square lattices that enable a systematic size scaling. As seen in the left panel of Fig.~\ref{fig:scaling}, we find the average energy gap $\Delta$ between the optimal and second-best solution at the end of the sweep to close only polynomially with system size, with an exponent close to $0.5$. 
At nonvanishing strength of the driver Hamiltonian $H_\mathrm{in}$, these two classical configurations will acquire energy shifts and encounter a level crossing, which is avoided by coupling the configurations via $H_\mathrm{in}$. The relevant order for this coupling in perturbation theory is given by the Hamming distance $d_\mathrm{H}$, i.e., by the number of classical bits by which the two lowest solutions differ \cite{Altshuler2010}. 
As seen in the right panel of Fig.~\ref{fig:scaling}, the Hamming distance increases only logarithmically with system size, i.e., the two lowest solutions will be coupled in low order. 
These findings suggest an avoided crossing whose gap closes only polynomially or quasi-polynomially with system size, leading us to expect the quantum annealing to perform efficiently. 
A number of straightforward improvements, such as higher-order driving terms, can further enhance the efficiency of the quantum algorithm \cite{Hauke2020}. 

Further, since our problem requires only an undirected solution of the shortest path,  one could e.g.\ adapt the algorithm developed in Ref.~\cite{Bauckhage2018}. 
There, the qubits encode the vertices rather than the links. The final state of the quantum computer then returns the vertices that lie on the optimal path, but not their order, which thus requires additional postprocessing. If there is a unique shortest path, the number of qubits required for the algorithm of Ref.~\cite{Bauckhage2018} scales as $|\mathcal{V}|$. In the presence of several shortest paths, the number of links in the shortest path $\Delta$ needs to be known and the scaling grows to $|\mathcal{V}|\Delta$. 
In contrast, the qubit requirement of our algorithm scales slower than $2|\mathcal{V}|c$, where $c$ denotes the number of edges at the site with the largest connectivity, and no \textit{a priori} knowledge of properties of the shortest path is necessary.

Let us now discuss the first illustrative application of our quantum computing scheme, which we have performed using the D-WAVE  facility.  To keep the computational effort as low as possible, we considered transitions in a smooth two-dimensional double well potential (shown in the right panel of Fig.~1 and defined in the SM \cite{SM}). For such a simple system, only 20 random configurations are sufficient for a reasonable coarse-grained representation of configuration space. To construct the associated  graph,  the nearest neighbors of each configuration were selected according to cut-off distance of 0.6, leading to 78 spins. The edges were computed from Eq. (5) in the low temperature limit (see SM \cite{SM}). We then considered the  generalized Ising Hamiltonian defined in Eqs.~(1) through (5) of the SM, with $A=100$ and $B=50$. 

Even for such a simple system, computing the dominant path by a simulated annealing algorithm on a classical computer required more than one hour of calculation on a standard desktop computer. However, the same optimization problem was solved in 3 seconds (on average) when resorting to a hybrid scheme combining classical and quantum annealing, with a total usage of  about 400 ms of quantum computing time. (Details about this hybrid scheme are given in the D-WAVE documentation, which was provided upon request to  D-WAVE Systems, Inc.).  The results are shown in the right panel of Fig.~1.  We note that we were not able to solve the optimization problem by resorting exclusively to quantum annealing on the current machine. This finding suggests that---at least in the current era of noisy intermediate-scale quantum devices---a hybrid classical/quantum annealing approach may be particularly efficient for this type of problems.

What kind of quantum computing resources would be required for a realistic molecular simulation? Based on the illustrative application to protein folding discussed above, we expect the minimal number of qubits required to study a biologically relevant transition to be on the order of $10^5$. 
To date, the largest quantum annealing machines have $o(10^3)$ qubits and are not yet in a position to outperform classical  computers in  shortest path calculation. However, if the current exponential growth in quantum processors continues, the proposed method has the potential to play a game changing role in the field of molecular simulations, enabling the investigation of  increasingly complex and rare transitions.

Thus, while immediate practical applications are still limited, our approach represents a new paradigm for tackling biophysical problems using quantum optimization:  
Instead of discretizing real space onto a lattice ---which would hamper the application to realistic systems of biological interest--- it makes efficient use of the quantized nature of the qubit register to represent a discrete set of system's configurations, each of which can be arbitrarily complex. In addition, it can capitalize on recently developed  powerful machine learning techniques to perform an uncharted exploration of complex energy landscapes. 

\textit{Acknowledgements.---}
We thank R.\ Covino and A.\ Laio for important discussions,  D.E.S. Research for providing their MD simulations, and D-WAVE for granting free access to their quantum annealing machine. 
P.H.\ acknowledges support by Provincia Autonoma di Trento and the ERC Starting Grant StrEnQTh (Project-ID 804305). 
This work was conceived within Q@TN---Quantum Science and Technology in Trento.

\appendix

\renewcommand{\theequation}{S\arabic{equation}}%labels equations with Eq. (S1) etc. 
\setcounter{equation}{0}%sets counter for the equations to 1, so that start with (S1)
\renewcommand{\thefigure}{S\arabic{figure}}%the same for the figures
\setcounter{figure}{0}

\onecolumngrid

\vspace*{0.75cm} 
\begin{center}
{\Large{ \textbf{Supplementary Material to\\\vspace*{0.2cm} \textit{Dominant Reaction Pathways by Quantum Computing}}}}
\end{center}
\vspace*{0.1cm} 
This supplementary material reports the explicit derivation of the generalized Ising model onto which the shortest-path problem is mapped. This includes modified constraints that enable to avoid loops in the shortest path.  We also provide details about  illustrative applications of our classical and quantum schemes on low-dimensional energy surfaces, the DRP Calculations, and the finite-size scaling of the final gap and Hamming distance. 
\vspace*{0.75cm} 

\twocolumngrid

\section*{Generalized Ising Hamiltonian} 
The generalized Ising Spin formulation of the shortest-path problem on the discrete graph is straightforwardly obtained from the quadratic function  (7) of the main text by means of the substitution  $x_{(ij)}= (\sigma^z_{(ij)}+1)/2$. The result is 
\be
\label{eq:Ising_HamiltonianA}
H_A &=& H_{(\mathrm{s})}+ H_{(\mathrm{t})}+ H_{(ij)}\\
H_{(\mathrm{s})}& =&\frac{A}{4}\left\{4+\sum_{i,j \in \mathcal{V}}( \sigma^z_{(\mathrm{s}j)}\sigma^z_{(\mathrm{s}i)}+\sigma^z_{(j\mathrm{s})}\sigma^z_{(i\mathrm{s})} -\sigma^z_{(j\mathrm{s})}\sigma^z_{(\mathrm{s}i)}\right.\nonumber \\
&&-\left. \sigma^z_{(\mathrm{s}j)}\sigma^z_{(i\mathrm{s})})  - 4 \sum_{i\in \mathcal{V}}\sigma^z_{(\mathrm{s}i)} + 4 \sum_{i\in \mathcal{V}}\sigma^z_{(i\mathrm{s})}\right\}\\
H_{(\mathrm{t})}& =&\frac{A}{4}\left\{4+\sum_{i,j \in \mathcal{V}}( \sigma^z_{(\mathrm{t}j)}\sigma^z_{(\mathrm{t}i)}+\sigma^z_{(j\mathrm{t})}\sigma^z_{(i\mathrm{t})} -\sigma^z_{(j\mathrm{t})}\sigma^z_{(\mathrm{t}i)}\right.\nonumber \\
&&-\left. \sigma^z_{(\mathrm{t}j)}\sigma^z_{(i\mathrm{t})})  +4 \sum_{i\in \mathcal{V}}\sigma^z_{(\mathrm{t}i)} -4 \sum_{i\in \mathcal{V}}\sigma^z_{(i\mathrm{t})}\right\}\\
H_{(ij)} &=&\frac{A}{4}\left\{\sum_{ \stackrel{i, j, k \in \mathcal{V}}{i\ne \mathrm{s}, \mathrm{t}}  } (\sigma^z_{(ij)}\sigma^z_{(ik)}+\sigma^z_{(ji)}\sigma^z_{(ki)} - \sigma^z_{(ji)}\sigma^z_{(ik)} \right.\nonumber\\
&& \left. - \sigma^z_{(ij)}\sigma^z_{(ki)})\right\}\,
\ee
while, dropping an irrelevant constant in $H_B$, we get 
\be
\label{eq:Ising_HamiltonianB}
H_B = \frac{B}{2} \sum_{(ij) \in \mathcal{E}} w_{(ij)} \sigma_{(ij)}^z\,.
\ee
The task is now to find the ground state of the qubit (Ising-spin) Hamiltonian $H=H_A+H_B$.

\begin{figure}[t!]
	\includegraphics[width=.5\textwidth]{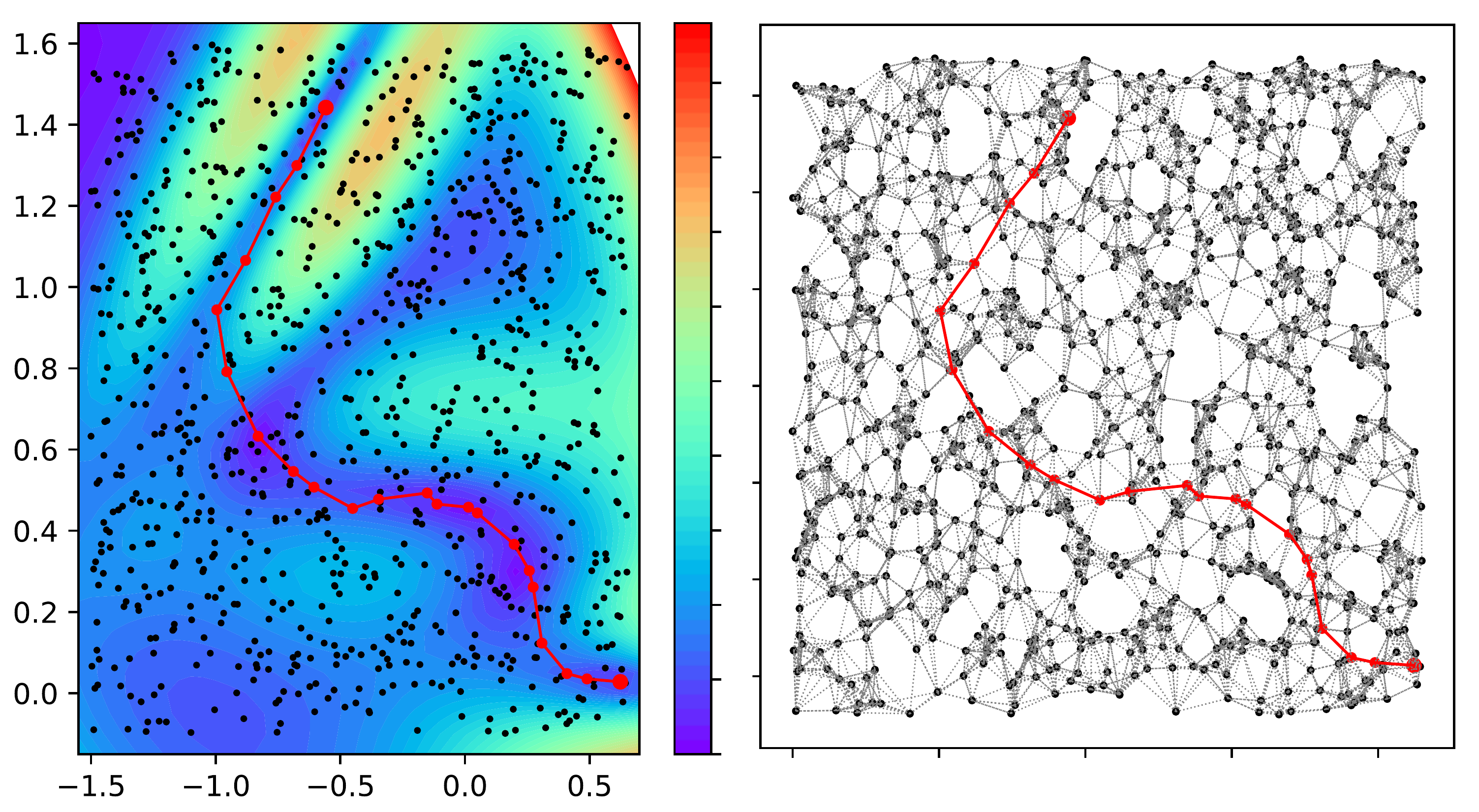}
	\caption{Thermally activated transitions in the M\"uller Brown energy surface. In the left panel, the heat-map displays the potential energy in units of thermal energy $k_BT$,  while the points have been sampled from a flat distribution. The red line is the the minimum-energy path calculated using the Dijkstra algorithm.   The right panel shows the associated weighted graph and (in red) the corresponding optimal path.}
	\label{fig:concept}
\end{figure}

In quantum annealing \cite{Das2005,Das2008,Albash2018,Hauke2020}, this problem is solved by initializing the qubits in the ground state of an easily solvable Hamiltonian that does not commute with $H$, say $H_\mathrm{in}=-h\sum_{(ij)}\sigma_{(ij)}^x$. Then, the Hamiltonian $H_\mathrm{qb}(t)$ of the qubit system is slowly deformed over time, $H_\mathrm{qb}(t)=a(t) H_\mathrm{in} + b(t) H$, with scheduling functions $a(t)$ and $b(t)$. The initial condition is $a(0)=1$ and $b(0)=0$, while the sweep ends at $t_\mathrm{sweep}$ with $a(t_\mathrm{sweep})=0$ and $b(t_\mathrm{sweep})=1$. In this way, the Hamiltonian is transformed from the simple $H_\mathrm{in}$ to the complex problem Hamiltonian $H$. The adiabatic theorem implies that if the sweep is performed sufficiently slowly as compared to the minimal energy gap encountered, the system remains in its instantaneous ground state. The final state at the end of the protocol is the ground-state of $H$, i.e.\ the solution to the shortest-path problem. The solution can then be simply read out by measuring the qubits in the computational basis of $\sigma^z$ eigenvalues.

In practice, the shortest path will avoid loops, and the constraints in Eq.~(6) of the main text then amount to
\begin{align}
&\forall i: \left(\sum_{j\in \mathcal{V}} x_{(ij)},\sum_{j\in \mathcal{V}} x_{(ji)}\right) = \begin{cases}(1,0), &\text{if }i=\mathrm{s};\\ (0,1), &\text{if }i=\mathrm{t};\\ (0,0)\,\, \mathrm{or} \,\, (1,1), &\text{else.}\end{cases}
\end{align}
One may implement this observation with an additional set of quadratic terms such as $\sum_{ \stackrel{i \in \mathcal{V}}{i\ne \mathrm{s}, \mathrm{t}}}(\sum_{j\in \mathcal{V}} x_{(ij)}-1)^2$ etc., which one can add to Eq.~\eqref{eq:Ising_HamiltonianA}. In addition, the edges leading into $\mathrm{s}$ and out of $\mathrm{t}$ can be eliminated. We find the more flexible formulation in Eq.~(6) of the main text that allows for loops to be sufficient for our purposes.

\section*{Transition paths in two-dimensional toy models.}
To illustrate  our classical and quantum computing algorithms for transition  path calculation, here we have investigated  thermally activated transitions undergone by a point particle diffusing in two 2-dimensional prototypical toy models.  

{\bf Transition path calculation by classical computing}. We begin by considering  the so-called Muller-Brown \cite{Muller} potential, a two-dimensional  energy surface shown in  Fig. S1 and defined by  
\be
U(x,y) = \sum_{i=1}^4 A_i e^{a_i(x-x^i_0)^2 + b_i(x-x_0^i) (y-y_0^i) + c_i (y-y_0^i)^2}\,,
\ee
where the parameters are reported in the Table~I. 

Our goal is to find the dominant reaction path connecting  the local energy minimum $Q_R=-(0.56, 1.44)$ to the  energy minimum at $Q_P=(0.62, 0.03)$. 
To facilitate the visual assessment of the results, we  focus on the low-temperature regime, in which the SH functional reduces to 
\be
S_{\mathrm{HJ}}= (k_BT/2)^{-1}\int_{Q_i}^{Q_f} dl |\nabla U[Q(l)]|.
\ee
Thus, in this limit, the dominant pathway reduces to the minimum-energy path \cite{DRP0}. 
As a first step,  we generated an ensemble of 2000 configurations by randomly sampling points in the region $q_1 \in [-1.5, 0.65], q_2\in [0.65, 1.6]$. In  applications to macromolecular systems, this random ensemble of points would be replaced by molecular conformations generated on a classical machine.  We emphasize that the flat distribution of points on the plain does not correspond to any physically meaningful ensemble, though it does provide a practical testbed.  
Next, we built the sparse network connecting each point to its 10 nearest neighbors and assigned a weight to each edge proportional to the corresponding contribution $w_{(ij)}$ to the HJ action (inset in the left panel of Fig.\ S1).  Finally, to compute the shortest path on this graph we resorted to a classical computer, using  the Dijkstra algorithm.  The result is the red line, which clearly provides a very accurate representation of the minimum-energy path in this landscape.
\begin{table}[htp]
	\caption{Parameters of the M\"uller-Brown potential}
	\begin{center}
		\begin{tabular}{|c|c|c|c|}
			$A_1$=-200, & $A_2$=-100, & $A_3$=--170, & $A_4$=15\\
			\hline
			$x_0^1$=1, &  $x_0^2$=0, &  $x_0^3$=-0.5,&  $x_0^4$=1\\
			\hline
			$ y_0^1$=0, & $y_0^2$=0.5, &  $y_0^3$=-1.5, & $y_0^4$=1\\
			\hline
			$ a_1$=1, & $a_2$=-1,&  $a_3$=--6.5,&  $a_4$=0.7\\  
			\hline
			$ b_1$=0,& $b_2$=0,&  $b_3$=-11,&  $b_4$=0.6,\\
			\hline
			$c_1$=-10, & $c_2$=-10,&  $c_3$=--6.5,&  $c_4$=0.7.
		\end{tabular}
	\end{center}
\end{table}

{\bf Transition path sampling by quantum annealing.}
In the main text, we discuss a calculation of the minimum energy path in the two-dimensional double well potential,
\be
U(x,y) = u_0 (x^2-x_0^2)^2 + \frac{1}{2}k_0 y^2,
\ee
where $u_0=k_BT=1$, and $k_0=2$ and $x_0=1$, in appropriate units. To this goal,  we computed the global minimum of the generalized Ising Hamiltonian, by combining the classical simulating annealing and quantum annealing protocols available in the OCEAN suite, operating on the D-WAVE quantum annealing machine.

\section*{Details on the DRP Calculation of Folding Pathways of Protein Trp-Cage}

{\bf Generation of the initial and final configurations.}
20 initial unfolded conformations were randomly retrieved from the plain MD trajectories made available by D.E.S. Research by
randomly sampling states with fraction of native contacts (Q) lower than 0.1 that are separated by at least
one folding-unfolding event.

Protein topologies were generated using Charmm22 force field
with TIPS3P water model. Lys, Arg, Asp, and Glu residues as well as the N-termini and C-termini were treated in their charged states. Each initial condition was solvated in a
cubic box with size 20 \AA ~greater than the maximum diameter of the system and neutralized
with the appropriate number of Na$^+$ and Cl$^-$ ions. Each system was then energy minimized
using the steepest descent algorithm then equilibrated in a 500 ps NVT simulation carried
out by restraining the protein heavy atoms using a harmonic potential with constant 1000 kJ~mol$^{-1}$nm$^{-2}$.

{\bf Generation of molecular configurations in the transition region.}   
To produce an ensemble of molecular configurations (to be associated to vertices of our sparse graph), we adopted the following procedure.  From each of 20 initial unfolded configurations,  we used 3 iterations of the Self-Consistent Path Sampling (SCPS) algorithm \cite{SCPS} to generate 30 folding trajectories, reaching the native state in 15$\times 10^5$ integration steps.  Configurations were saved every 500 steps and configurations with a fraction of native contact $Q>0.9$ were discarted.  Further details about the SCPS implementation and the parameters used can be found in Ref.~\cite{HET-s}. 

{\bf Graph Construction and Cost Calculation.} The ensemble of configurations generated by SCPS was downsampled to 2180 independent configurations. We assigned edges in the graph to pairs of  vertices $i$ and $j$  separated by a Root-Mean-Square (RMSD) distance $\Delta l_{(ij)}<3$\AA.  According to the DRP theory described in the main text, the  cost of each edge  $w_{(ij)}$ is calculated according to 
\be\label{SHJ1}
w_{(ij)} &=&  \frac{ \Delta l_{(ij)}}{2 \sqrt{D}} \left( L_i+L_j \right),\\
L_i &=& \sqrt{V_{\mathrm{eff}}(Q_i)+E_{\mathrm{eff}}}.
\ee
In this equation, $Q_i$ is the molecular conformation (i.e., the set of $3N$ atomic coordinates) associated with the $i-$th vertex,  
\be\label{V}
V_{{\mathrm{eff}}}(Q)= \sum_{i} \frac{1}{4 \gamma_i m_i k_BT} \left( \nabla_i U(Q)^2 - 2 k_BT \nabla^2_i U(Q)\right)\nonumber\\
\ee
is called the effective potential, and $U(Q)$ is the molecular potential energy calculated in the configuration $Q$.  The effective energy $E_{\textrm{eff}}$ is a free parameter which controls the total time of the transition (see discussion in the main text). 

In principle, the effective potential at each configuration $Q$ may be explicitly calculated  from the  force field. This calculation is quite expensive, since the second term in Eq. (\ref{V}) requires to compute  numerically the divergence of the total  force acting on each atom.  In this first exploratory calculation, we chose to take a much simpler approach and estimated the effective potential within a mean-field approximation. To derive such a scheme, we recall  that 
at mean-field level,  the time $\Delta t$  taken on average to diffuse from $Q$ for an infinitesimal distance $\Delta l$  along a transition path is given by \cite{DRP2}
\be
\Delta t(Q) =  \frac{\Delta l}{\sqrt{4 D (E_{\textrm{eff}}+ V_{\textrm{eff}}(Q)}}\,.
\ee
Therefore, the  integrand $\sqrt{4D (V_{\textrm{eff}}(Q_i) + E_{\textrm{eff}})}$  in the Hamilton--Jacobi functional  
may be identified with the  total  mean-square velocity of all the atoms in the system. 
Since the Langevin equations used to derive the DRP theory describe the dynamics of the system   coupled  to a thermostat, the mean-square velocity can be considered constant. Thus, at mean field level, the cost function 
$w_{(ij)}$ is simply given by the RMSD distance $ \Delta l_{ij}$  up to an irrelevant normalization factor.

\section*{Efficiency of quantum annealer}

In the main text, we have mapped the task of finding the dominant reaction pathways to a shortest-path optimization problem. It is known that these problems can be solved in polynomial time on a classical computer \cite{Fredman1987,Dijkstra}. Here, we address the question how efficient a quantum annealer may be in solving these problems.

{\bf Polynomial scaling of final gap.}
To assess the complexity scaling of the quantum annealer in the proposed optimization problem, we first characterize the energy landscape of the Ising optimization problem given in Eqs.~(7) and~(8) of the main text, respectively Eqs.~\eqref{eq:Ising_HamiltonianA} and Eqs.~\eqref{eq:Ising_HamiltonianB}. As the efficiency of quantum annealers is primarily determined by the structure of gap closings \cite{Das2005,Das2008,Albash2018,Hauke2020}, we first focus on the ground-state gap $\Delta$ at the end of the sweep, i.e., the energy difference between the optimal and the second best solution. 

To this end, we exploit the efficiency of solving the classical shortest path problem using linear programming. To facilitate finite-size scaling, we take two exemplary regular lattices, a square lattice with edges between nearest neighbors (NNs) and one with edges up to next-to-next-to-nearest neighbors (NNNNs). We assume $A>B|\mathcal{E}|$, enabling us to hard code the constraints into the solution. 

\begin{figure}
	\includegraphics[width=.45\textwidth]{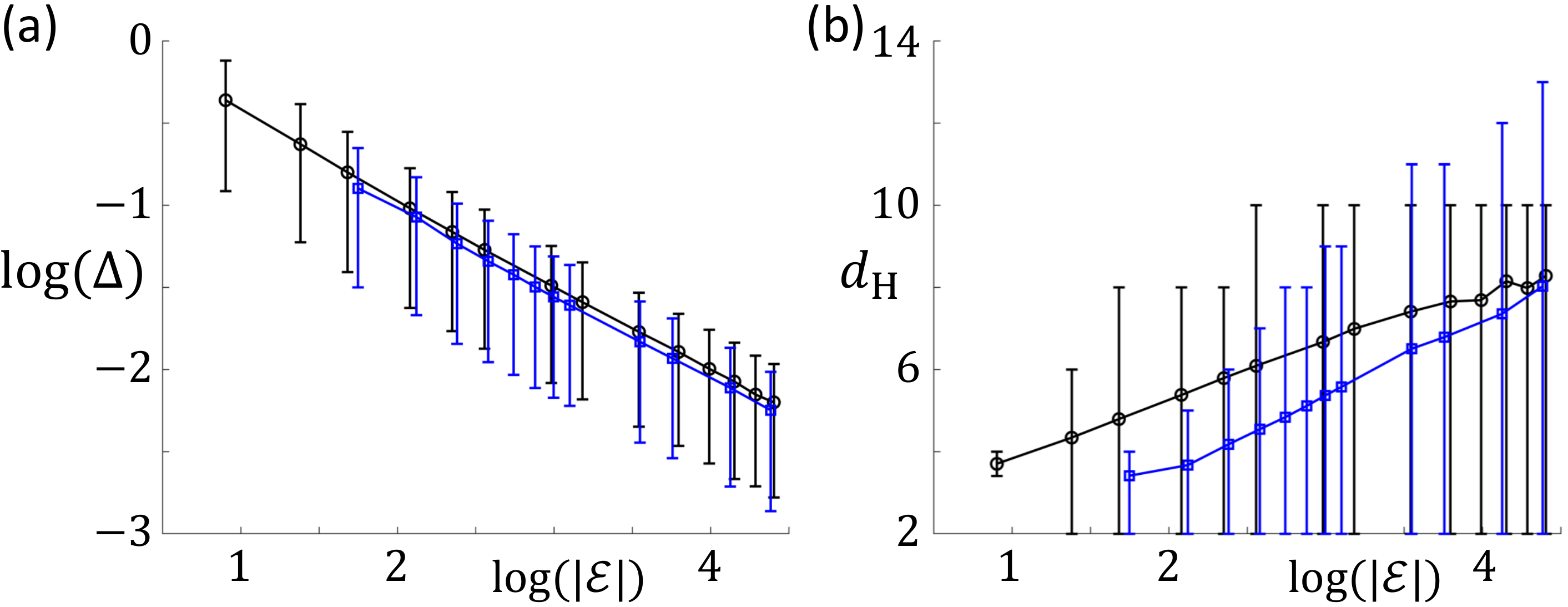}
	\caption{
		Scaling of average gap and Hamming distance between optimal and second-best solution versus number of edges. 
		Data is for the shortest path problem on a square lattice, with edges along nearest neighbors (NNs, black lines and circles) and up to next-to-next-to-nearest neighbors (NNNNs, blue lines and squares). 
		Error bars are $67\%$ confidence intervals and dashed lines are linear fits. 	
		(a) Log-log plot, revealing the approximate polynomial scaling of the average final gap as a function of number of edges. The curves for different lattices approximately collapse. 
		(b) Log-linear plot, revealing the approximate logarithmic scaling of the average Hamming distance.
	}
	\label{fig:scaling_vs_edges}
\end{figure}

\begin{figure}
	\includegraphics[width=.45\textwidth]{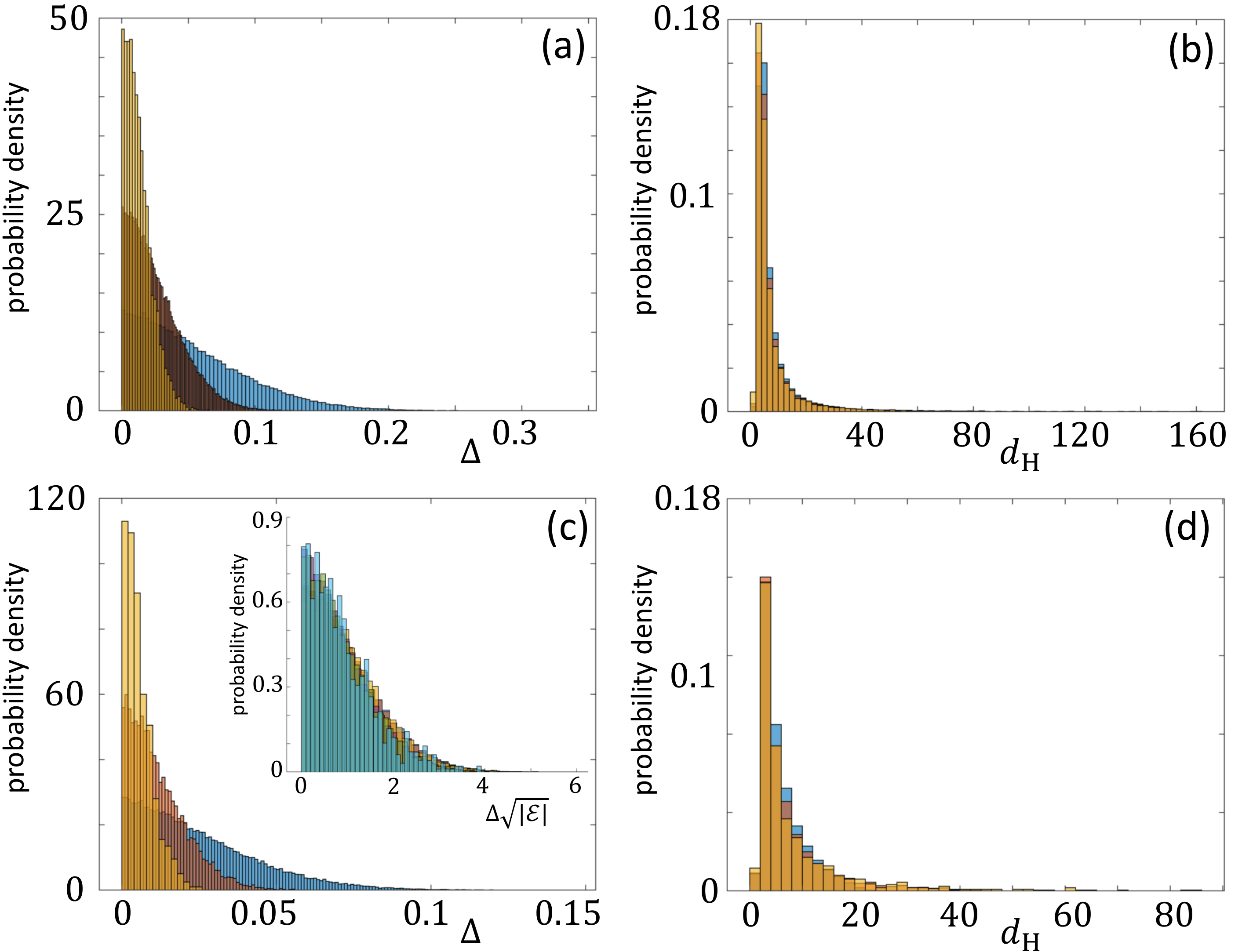}
	\caption{
		Histograms of final gap $\Delta$ (left column) and Hamming distance $d_\mathrm{H}$ (right column) for lattice sizes $|\mathcal{V}|=10\times 10$, $20\times 20$, and $40\times 40$. 
		Upper row: NN lattice. Lower row: NNNN lattice. 
		(a,c) The gap distribution shifts towards smaller values as the lattice size increases. Inset in panel c: The histograms of panels a and c shows a good collapse when rescaled as $\Delta \sqrt{|\mathcal{E}|}$, further corroborating the polynomial scaling of the final gap, not only in average, but in distribution. 
		(b,d) The strong peak at small $d_\mathrm{H}$ indicates the two best solutions typically differ only by few qubits. The distribution shifts only slowly with increasing lattice size and the tail at large $d_\mathrm{H}$ has only a rather small weight. 
	}
	\label{fig:histograms}
\end{figure}

In the left panel of Fig.~2 of the main text, we plot the gap for these lattices, averaged over many realizations (between $10^5$ and $10^3$), with weights $w_{(ij)}$ drawn uniformly from the interval $[0,1]$. 
Within the system sizes accessible to our numerics ($|\mathcal{V}|=80\times 80$ for NN and $|\mathcal{V}|=40\times 40$ for NNNN lattice) the gap decreases polynomially with system size. 
We extract the exponent from a simple fit using the function $\Delta= c_1|\mathcal{V}|^{c_2}$, omitting the few smallest system sizes. 
We obtain the exponents $c_2=-0.52$ (NN lattice) and $-0.55$ (NNNN lattice), which are approximately independent of the lattice coordination number (i.e., the ratio of edges over vertices). For completeness, the fitted prefactors are $c_1=0.80$ and $0.61$, respectively. 
In Fig.~\ref{fig:scaling_vs_edges}a, we plot the same data against $|\mathcal{E}|$, indicating that the average value of the gap scales polynomially with the number of edges, which equals the number of required qubits. 
The histograms in Fig.~\ref{fig:histograms}a,c indicate that this scaling holds not only for the average value, but for the entire gap distribution. As the inset in Fig.~\ref{fig:histograms}c shows, we obtain a good collapse of the gap distributions when rescaled as $\Delta \sqrt{|\mathcal{E}|}$.

{\bf Logarithmic scaling of final Hamming distance.}
These findings suggest the potential for a polynomial efficiency of the quantum annealer. However, the adiabatic theorem is concerned with the \textit{minimal} gap during the sweep, not only the final gap. To access this quantity in a full quantum computation is, however, extremely challenging due to the exponential scaling of the quantum mechanical Hilbert space, and can be done only for rather small systems \cite{Caneva2007,Hauke2015} or for specific models such as the p-spin model \cite{Susa2018}. 

Nevertheless, one can obtain good estimates on the minimal gap by perturbative arguments starting from the classical target problem \cite{Altshuler2010}. At non-vanishing strength of the transverse-field driver Hamiltonian $H_\mathrm{in}=-h\sum_{(ij)}\sigma_{(ij)}^x$, the low-lying levels will acquire energy shifts leading them to intersect. Exact crossings will be avoided as $H_\mathrm{in}$ couples the energy levels, usually in some high order of perturbation theory. The relevant perturbative order is given by the Hamming distance $d_\mathrm{H}$ between two classical solutions, i.e., the number of qubits that $H_\mathrm{in}$ needs to flip to proceed from one configuration to the other. 

As Fig.~\ref{fig:scaling_vs_edges}b as well as the right panel of Fig.~2 of the main text indicate, the Hamming distance between the optimal and second best solution increases to good approximation logarithmically with system size. 
A fit as $d_\mathrm{H}= c_1'+\log(|\mathcal{V}|^{c_2'})$ yields $c_1'=3.6$, $c_2'=1.3$ (NN lattice) and $c_1'=1.3$, $c_2'=2.1$ (NNNN lattice). In contrast to the average final gap, the coordination number seems to enter the coefficients $c_{1,2}'$. 
More importantly, at least within the considered lattices, the Hamming distance between the best solution typically is rather small. Large Hamming distances appear only in few instances, see the histograms in Fig.~\ref{fig:histograms}b,d. 

These results indicate that the avoided crossing between the lowest solutions typically appears in low perturbative order in $H_\mathrm{in}$. Together with the polynomially decreasing energy difference at the end of the sweep, this finding suggests a benign gap that is only polynomially or quasi-polynomially small at the avoided crossing, rather than the exponentially small gap that is often found \cite{Caneva2007,Altshuler2010}. 
Similarly, the best known classical algorithms for the shortest path problem, such as Dijkstra's algorithm with Fibonacci heap, also scale polynomially \cite{Fredman1987,Dijkstra}. 
There are numerous straightforward ways for further improving the quantum algorithm \cite{Hauke2020}. A simple possibility, at least in principle, is given by more complex driver Hamiltonians such as the non-stoquastic $H_\mathrm{in}=-J_x\sum_{(ij),(k\ell)}\sigma_{(ij)}^x \sigma_{(k\ell)}^x$. Such a driver Hamiltonian requires less than $d_\mathrm{H}$ applications to couple two classical configurations, which reduces the relevant perturbative order at the avoided crossing and thus can lead to an even slower gap closing. 
Although our estimates are based on finite-precision numerical data and specific lattice structures, the shortest path problem thus seems a good candidate for studying a potential quantum speedup versus the best known classical algorithms. 
Importantly, although it is typical for quantum annealers to not change the complexity class of the problem \cite{Hen2011}, already a polynomial speedup could have huge practical implications, as is famously illustrated by the Grover search algorithm \cite{Grover1996}.


\begin{thebibliography}{99}
	\bibitem{enhanced}  Y.I. Yang, Q. Shao, J.  Zhang, L. Yang, and Y.Q.  Gao, J. Chem Phys. {\bf 151} 070902 (2019). 
	\bibitem{METAD} A. Laio, M. Parrinello, Proc.  Natl. Acad.  Sci. USA {\bf 99} 12562 (2002). 
	\bibitem{AMD} M. A. Cuendet and M. E. Tuckerman,  J. Chem. Theory Comput. {\bf 10}, 2975 (2014).
	\bibitem{BFA} S. A. Beccara, L. Fant, and P. Faccioli, Phys. Rev. Lett. {\bf 114}, 098103 (2015). 
	\bibitem{TPS} C. Dellago, P.G. Bolhuis, F.S. Csajka, D. Chandler, J. Chem. Phys. {\bf 108} (5), 1964 (1998). 
	\bibitem{MSM} B. E. Husic,  V. S. Pande, J. Am. Chem. Soc. {\bf 140} 2386 (2018).
	\bibitem{milestoning}  A. K. Faradjian and R. Elber, J. Chem. Phys.  {\bf 120} 10880 (2004).
	\bibitem{Arute2019} F. Arute et al., Nature {\bf 574}, 505--510 (2019). 
	\bibitem{Das2005} A. Das and B.K. Chakrabarti, Lecture Notes in Physics book series, ISSN 0075-8450, Springer-Verlag Berlin/Heidelberg (2005). 
	\bibitem{Das2008} A. Das and B.K. Chakrabarti, Rev. Mod. Phys. {\bf 80}, 1061 (2008). 
	\bibitem{Albash2018} T. Albash, D.A. Lidar, Rev. Mod. Phys. {\bf 90}, 015002 (2018). 
	\bibitem{Venegas2018} S.E. Venegas-Andraca, W. Cruz-Santos, C. McGeoch, M. Lanzagorta, Contemporary Physics
	{\bf 59}, 174–197 (2018). 
	\bibitem{Hauke2020} P. Hauke, H.G. Katzgraber, W. Lechner, H. Nishimori, W.D. Oliver, Rep. Prog. Phys.  {\bf 83}, 054401 (2020). 
	\bibitem{Streif2018} M. Streif, F. Neukart, M. Leib, In: Feld S., Linnhoff-Popien C. (eds) Quantum Technology and Optimization Problems. QTOP 2019. Lecture Notes in Computer Science, vol 11413. Springer, Cham (2018). 
	\bibitem{Li2018} R.Y. Li, R.D. Felice, R. Rohs, D.A. Lidar, npj Quantum Information {\bf 4}, 14 (2018). 
	\bibitem{Genin2019} S.N. Genin, I.G. Ryabinkin, A.F. Izmaylov, 	arXiv:1901.04715 (2019). 
	\bibitem{Cao2019} Y. Cao et al., Chem. Rev. {\bf 119}, 10856–10915 (2019). 
	\bibitem{Matsuura2020} S. Matsuura, T. Yamazaki, L. Huntington, A. Zaribafiyan, New J. Phys. {\bf 22}, 053023 (2020). 
	\bibitem{Outeiral2020} C. Outeiral, M. Strahm, J. Shi, G.M. Morris, S.C. Benjamin, C.M. Deane, WIREs Comput. Mol. Sci. e1481 (2020). 
	\bibitem{QMD1} A. Perdomo-Ortiz, N.  Dickson, M. Drew-Brook,  G. Rose, and A. Aspuru-Guzik, Scientific Reports {\bf 2}, 571 (2012).
	\bibitem{QMD2} L.H. Lu  and Y.Q. Li  Chin. Chem.  Lett. {\bf 36}, 080305 (2019).
	\bibitem{lattice_models} L. Mirny and E.  Shakhnovich  Ann. Rev. Biophys. Biomol. Struct. {\bf 30} 361 (2001). 
	\bibitem{Covino} E. Chiavazzo, R. Covino, R., Coifman, C. W. Gear, A. S.  Georgiou, G.  Hummer, and I. G. Kevrekidis,   Proc. Nat. Acad. Sci. U.S.A. {\bf 114} (28), E5494 (2017).
	\bibitem{DRP1} P. Faccioli, M. Sega, F. Pederiva, and H. Orland, Phys. Rev. Lett. {\bf 97}, 108101 (2006).
	\bibitem{DRP0} R. Elber and D. Shalloway, J. Chem. Phys. {\bf 112}, 5539 (2000).
	\bibitem{DRP2} M. Sega, P. Faccioli, F. Pederiva, G. Garberoglio, and H. Orland, Phys. Rev. Lett. {\bf 99}, 118102 (2007).
	\bibitem{hMD} P. Faccioli and F. Pederiva , Phys. Rev. {\bf E 86}, 061916 (2012)
	\bibitem{SM} See Supplemental Material for details about the generalized Ising model, illustrative applications on low-dimensional energy surfaces, the DRP Calculations, and the finite-size scalings, which includes Refs.~\cite{Muller,Fredman1987,Caneva2007,Hauke2015,Susa2018,Hen2011,Grover1996}.
	%
	\bibitem{Muller} K. M\"uller and L. D. Brown, Theor. Chem. Acta (Berl.) {\bf 53}, 75 (1979).
	\bibitem{Fredman1987}  M. L. Fredman, R. E. Tarjan, Journal of the ACM  {\bf 34 }, 596–615 (1987).
	\bibitem{Caneva2007}  T. Caneva, R. Fazio, G.E. Santoro, Phys. Rev. B {\bf 76}, 144427 (2007).
	\bibitem{Hauke2015}  P. Hauke, L. Bonnes, M. Heyl, W. Lechner, Front. Phys. {\bf 3 }, 21 (2015).
	\bibitem{Susa2018}  Y. Susa, Y. Yamashiro, M. Yamamoto, I. Hen, D. A. Lidar, H. Nishimori, Phys. Rev. A  {\bf 98 }, 042326 (2018).
	\bibitem{Hen2011}  I. Hen, A. P. Young, Phys. Rev. E  {\bf 84 }, 061152 (2011).
	\bibitem{Grover1996}  L. K. Grover, Proc. of the 28th annual ACM Symp on Theory of Computing. New York, USA: ACM Press, 6: 212-219 (1996). 
	%
	\bibitem{howfastfoldingproteinsfold} K. Lindorff-Larsen, S. Piana, R. O. Dror, and D. E. Shaw, Science {\bf 334} 517 (2011).
	\bibitem{SCPS} S. Orioli, S. A. Beccara, P. Faccioli, J. Chem. Phys. {\bf 147}, 064108 (2017).
	\bibitem{HET-s} L. Terruzzi, G. Spagnolli, A. Boldrini, J. R. Raquena, E. Biasini and P. Faccioli, PLoS Comp. Biol. 16(9): e1007922 (2020). 
	\bibitem{Dijkstra}  M. Sniedovich, Control and Cybernetics. {\bf 35}, 599 (2006).
	\bibitem{runningtime} T. H. Cormen, C.E. Leiserson, L. R. Rivest, and C. Stein, {\it Introduction to Algorithms (2nd ed.)},  MIT Press and McGraw–Hill,  595. ISBN 0-262-03293-7
	\bibitem{Altshuler2010} B. Altshuler, H. Krovi, and J. Roland, PNAS {\bf 107}, 12446-12450 (2010). 
	\bibitem{Bauckhage2018}	C. Bauckhage, E. Brito, K. Cvejoski, C. Ojeda, J. Sch\"ucker, R. Sifa, In Proceedings
	of 14th Int. Workshop Mining and Learning with Graphs (MLG’18). ACM, New York, NY, USA (2018). 
	
\end{thebibliography}
\end{document}